\documentclass[preliminary,copyright,creativecommons]{eptcs}
\providecommand{\event}{Proceedings QAPL 2011}
\usepackage{latexsym}
\usepackage{times}
\usepackage{amsmath}
\usepackage{amssymb}
\usepackage{stmaryrd}
\SetSymbolFont{stmry}{bold}{U}{stmry}{m}{n}
\usepackage{xspace}
\usepackage{enumerate}
\usepackage{graphicx}
\usepackage{breakurl}
\newtheorem{defi}{Definition}[section]
\newtheorem{theo}[defi]{Theorem}
\newtheorem{prop}[defi]{Proposition}
\newtheorem{lemm}[defi]{Lemma}
\newtheorem{coro}[defi]{Corollary}
\newtheorem{exam}[defi]{Example}
\newtheorem{rema}[defi]{Remark}

\newenvironment{definition}{\begin{defi} \rm}{\end{defi}}
\newenvironment{theorem}{\begin{theo} \rm}{\end{theo}}
\newenvironment{proposition}{\begin{prop} \rm}{\end{prop}}
\newenvironment{lemma}{\begin{lemm} \rm}{\end{lemm}}
\newenvironment{corollary}{\begin{coro}\rm}{\end{coro}}
\newenvironment{example}{\begin{exam}\rm}{ \hfill $\Box$\end{exam}}

\newfont{\bbb}{bbm10}       

\newcommand{\inp}{\mathop\in}                           
\newcommand{\Times}{\mathord{\kern.08em\cdot\kern.08em}}
\newcommand{\Defs}{\mathrel{:=}}
\newcommand{\plat}[1]{\raisebox{0pt}[0pt][0pt]{#1}}     
\newenvironment{itemise}{\begin{list}{$\bullet$}{\leftmargin 20pt
                        \labelwidth\leftmargin\advance\labelwidth-\labelsep
                        \topsep 2pt \itemsep 0pt \parsep 0pt}}{\end{list}}

\renewcommand{\phi}{\varphi}
\renewcommand{\vec}{\overrightarrow}
\newcommand{\aRel}{\ensuremath{\mathrel{{\mathcal R }}}}
\newcommand{\size}[1]{\mathrel{|{#1}|}}
\newcommand\Thm[1] {Thm.~\ref{thm:#1}}
\newcommand\Wide[1] {~~~#1~~~}
\newcommand\WideRM[1] {~~~\textrm{#1}~~~}
\newcommand\WIDERM[1] {\Wide{\WideRM{#1}}}
\def \cV{{\cal V}}

\newcommand{\Par}[1]{\|}
\newcommand{\parT}{\Par{\Act}}

\newcommand{\Act}{\ensuremath{\mathsf{Act}}\xspace}

\newcommand{\pCSP}{\ensuremath{\mathsf{pCSP}}\xspace}

\newcommand{\VV}{{\mathbb V}}

\newcommand{\ResultsN}{\ensuremath{{\mathbb V}}}

\newcommand{\Apply}{\ensuremath{{\mathcal A}^\text{d}}\xspace}
\newcommand{\RApply}{\ensuremath{{\mathcal A}}\xspace}

\newcommand{\Hleq}{\mathrel{\leq_{\rm Ho}}}
\newcommand{\Sleq}{\mathrel{\leq_{\rm Sm}}}

\newcommand{\pmMayleq}{\mathrel{{\sqsubseteq}_{\text{\rm pmay}}^{\raisebox{-.3em}{\scriptsize${\Omega}$}}}}
\newcommand{\pmMustleq}{\mathrel{{\sqsubseteq}_{\text{\rm pmust}}^{\raisebox{-.3em}{\scriptsize${\Omega}$}}}}

\newcommand{\pMustleq}{\mathrel{\relax{\sqsubseteq}_{\text{\rm pmust}}}}
\newcommand{\pMayleq}{\mathrel{\relax{\sqsubseteq}_{\text{\rm pmay}}}}

\newcommand{\failsime}{\mathrel{\lhd_{\raisebox{-.1em}{\tiny\it FS}}}}

\newcommand{\failsimpreo}{\sqsubseteq_{\it FS}}

\newcommand{\nrmMayleq}{\mathrel{{\sqsubseteq}_{\text{{\bf nr}may}}^{\raisebox{-.3em}{\scriptsize$\Omega$}}}}
\newcommand{\nrmMustleq}{\mathrel{{\sqsubseteq}_{\text{{\bf nr}must}}^{\raisebox{-.3em}{\scriptsize$\Omega$}}}}
\newcommand{\rrmMayleq}{\mathrel{{\sqsubseteq}_{\text{{\bf rr}may}}^{\raisebox{-.3em}{\scriptsize$\Omega$}}}}
\newcommand{\rrmMustleq}{\mathrel{{\sqsubseteq}_{\text{{\bf rr}must}}^{\raisebox{-.3em}{\scriptsize$\Omega$}}}}
\newcommand{\nrMayleq}{\mathrel{{\sqsubseteq}_{\text{{\bf nr}may}}}}
\newcommand{\nrMustleq}{\mathrel{{\sqsubseteq}_{\text{{\bf nr}must}}}}
\newcommand{\rrMayleq}{\mathrel{{\sqsubseteq}_{\text{{\bf rr}may}}}}
\newcommand{\rrMustleq}{\mathrel{{\sqsubseteq}_{\text{{\bf rr}must}}}}

\newcommand{\dist}[1]{\mathop{\mbox{$\mathcal D$}} ({#1})   } 
\newcommand{\pdist}[1]{\overline{#1}  } 
\newcommand{\subdist}[1]{\mathop{\mbox{$\cal D_{\textsl{sub}}$}} ({#1})   }
\newcommand{\support}[1]{\lceil{#1}\rceil}
\newcommand{\lift}[1]{\mathrel{\overline{#1}}}
\newcommand{\eDis}{\mathop{\varepsilon}}
\newcommand{\Exp}[1]{\textrm{Exp}_{#1}}
\newcommand{\ExpDeltaf}{\Exp{\Delta}(f)}
\newcommand{\Imgf}[1]{\textrm{Img}_f(#1)}
\newcommand{\Stop}{{\times}}

\newcommand{\setof}[2]{\{ \, #1 \, \mid \, #2 \, \}}
\newcommand{\sset}[1]{\{ {#1}  \}  } 
\def \Real{\mbox{\bbb R}}

\newcommand{\Rfun}{\$} 
\newcommand{\Equ}[1]{\buildrel{#1}\over{=}}
\newcommand{\match}{{^\natural}}
\makeatletter

\RequirePackage{ifthen}   
\newcommand{\ifemptyelse}[3]{\ifx\@@bullshit#1\@@bullshit#2\else#3\fi}
\newcommand{\ifnotempty}[2]{\ifx\@@bullshit#1\@@bullshit\else#2\fi}

\newbox\@topbox
\newbox\@botbox
\newcommand{\linfer}[3][{}]{%
    \setbox\@topbox\hbox{%
      \renewcommand{\arraystretch}{1}%
      $\begin{array}{l}#2\end{array}$}%
  \setbox\@botbox\hbox{%
    \def\tilde##1{\widetilde{##1}}%
    \renewcommand{\arraystretch}{1}%
    $\begin{array}{l}#3\end{array}$}%
  \ifdim\wd\@botbox<\wd\@topbox
    \setbox\@botbox\hbox to \wd\@topbox{\box\@botbox\hfil}
  \else
    \setbox\@topbox\hbox to \wd\@botbox{\box\@topbox\hfil}
  \fi
  \frac{\box\@topbox}{\box\@botbox}
  }
\newcommand{\linferSIDE}[4][{}]{%
    \setbox\@topbox\hbox{%
      \renewcommand{\arraystretch}{1}%
      $\begin{array}{lr@{}}#2&#4\end{array}$}%
  \setbox\@botbox\hbox{%
    \def\tilde##1{\widetilde{##1}}%
    \renewcommand{\arraystretch}{1}%
    $\begin{array}{l}#3\end{array}$}%
  \ifdim\wd\@botbox<\wd\@topbox
    \setbox\@botbox\hbox to \wd\@topbox{\box\@botbox\hfil}
  \else
    \setbox\@topbox\hbox to \wd\@botbox{\box\@topbox\hfil}
  \fi
  \frac{\box\@topbox}{\box\@botbox}
  }

\newcommand{\darE}[1]{\mathrel{\dar{#1}\kern-.3em\succ}}
\newcommand{\Move}{\rightarrow}
\def\mine{\@transition\mapstofill}
\def\goesto{\@transition\rightarrowfill}
\def\Goesto{\@transition\Rightarrowfill}
\def\ngoesto{\@transition\nrightarrowfill}
\def\nsgoesto{\@transition\nmapstofill}
\def\comesfrom{\@transition\leftarrowfill}
\def\nGoesto{\@transition\nRightarrowfill}
\def\@transition#1{\@ifnextchar[{\@@transition{#1}}{\@@transition{#1}[]}}
\newbox\@transbox
\newbox\@arrowbox
\def\mapstofill{$\m@th\mathord\mapstochar\mathord-\mkern-6mu%
  \cleaders\hbox{$\mkern-2mu\mathord-\mkern-2mu$}\hfill
  \mkern-6mu\mathord\rightarrow$}
\def\Rightarrowfill{$\m@th\mathord=\mkern-6mu%
  \cleaders\hbox{$\mkern-2mu\mathord=\mkern-2mu$}\hfill
  \mkern-6mu\mathord\Rightarrow$}
\def\nrightarrowfill{$\m@th\mathord-\mkern-6mu%
  \cleaders\hbox{$\mkern-2mu\mathord-\mkern-2mu$}\hfill
  \mkern-6mu\mathord\not\mathord-\mkern-6mu
  \cleaders\hbox{$\mkern-2mu\mathord-\mkern-2mu$}\hfill
  \mkern-6mu\mathord\rightarrow$}
\def\nmapstofill{$\m@th\mathord\mapstochar\mathord-\mkern-6mu%
  \cleaders\hbox{$\mkern-2mu\mathord-\mkern-2mu$}\hfill
  \mkern-6mu\mathord\not\mathord-\mkern-6mu
  \cleaders\hbox{$\mkern-2mu\mathord-\mkern-2mu$}\hfill
  \mkern-6mu\mathord\rightarrow$}
\def\nRightarrowfill{$\m@th\mathord=\mkern-6mu%
  \cleaders\hbox{$\mkern-2mu\mathord=\mkern-2mu$}\hfill
  \mkern-6mu\mathord\not\mathord=\mkern-6mu
  \cleaders\hbox{$\mkern-2mu\mathord=\mkern-2mu$}\hfill
  \mkern-6mu\mathord\Rightarrow$}
\def\@@transition#1[#2]%
   {\setbox\@transbox\hbox
       {\vrule height 1.5ex depth .7ex width 0ex\hskip0.25em$\scriptstyle#2$\hskip0.25em}
   \ifdim\wd\@transbox<1.5em
      \setbox\@transbox\hbox to 1.5em{\hfil\box\@transbox\hfil}\fi
   \setbox\@arrowbox\hbox to \wd\@transbox{#1}
   \ht\@arrowbox\z@\dp\@arrowbox\z@
   \setbox\@transbox\hbox{$\mathop{\box\@arrowbox}\limits^{\box\@transbox}$}
   \ht\@transbox\z@\dp\@transbox\z@
   \mathrel{\box\@transbox}}
\def\nxmapsto{\@transition\nmapstofill}
\def\xmapsto{\@transition\mapstofill}
\def\mapstofill{$\m@th\mathord\mapstochar\mathord-\mkern-6mu%
  \cleaders\hbox{$\mkern-2mu\mathord-\mkern-2mu$}\hfill
  \mkern-6mu\mathord\rightarrow$}
\def\nmapstofill{$\m@th\mathord\mapstochar\mathord-\mkern-6mu%
  \cleaders\hbox{$\mkern-2mu\mathord-\mkern-2mu$}\hfill
  \mkern-6mu\mathord\arrownot\mathord-\mkern-6mu
  \cleaders\hbox{$\mkern-2mu\mathord-\mkern-2mu$}\hfill
  \mkern-6mu\mathord\rightarrow$}
\newcommand{\ar}[1]{\mathrel{\goesto[{#1}]}}
\newcommand{\nar}[1]{\hspace{6pt}\not\hspace{-6pt}\mathrel{\goesto[{#1\;}]}}
\newcommand{\dar}[1]{\mathrel{\Goesto[\raisebox{.08em}{\scriptsize$#1$}]}} 
\newcommand{\Ref}[2]{#1 \nar{#2}}
\newcommand{\RefE}[2]{#1 \dar{}\Ref{}{#2}}

\begin{document}

\def\titlerunning{Real-Reward Testing for Probabilistic Processes}
\def\authorrunning{Y. Deng, R.J. van Glabbeek, M. Hennessy \& C.C. Morgan}
\title{\titlerunning\\
  \large   (Extended Abstract)}

\author{Yuxin Deng$^1$%
\thanks{Deng was supported by the National Natural Science Foundation of China (61033002).}
\quad Rob van Glabbeek$^2$
\quad Matthew Hennessy$^3$%
\thanks{Supported by SFI project SFI 06 IN.1 1898.}
\quad Carroll Morgan$^4$%
  \thanks{Morgan acknowledges the support of ARC Discovery Grant DP0879529.}
\institute{\makebox[1.7em][r]{$^1$} Shanghai Jiao Tong University and Chinese Academy of Sciences, China}
\institute{\makebox[1.7em][r]{$^{2}$} National ICT Australia, Australia}
\institute{\makebox[1.7em][r]{$^3$} Trinity College Dublin, Ireland}
\institute{\makebox[1.7em][r]{$^{2,4}$} University of New South Wales, Australia}
}

\maketitle

\begin{abstract}
We introduce a notion of real-valued reward testing for probabilistic
processes by extending the traditional nonnegative-reward testing with
negative rewards. In this richer testing framework, the may and must
preorders turn out to be inverses.  We show that for convergent
processes with finitely many states and transitions, but not in the
presence of divergence, the real-reward must-testing preorder
coincides with the nonnegative-reward must-testing preorder. To prove
this coincidence we characterise the usual resolution-based testing in
terms of the weak transitions of processes, without having to involve
policies, adversaries, schedulers, resolutions, or similar structures
that are external to the process under investigation. This requires
establishing the continuity of our function for calculating testing
outcomes.
\end{abstract}

\section{Introduction}\label{sec:intro}
Extending classical testing semantics \cite{DNH84,henn} to a
setting in which probability and nondeterminism co-exist was initiated
in \cite{WL92}.  The application of a test to a process yields a set
of probabilities for reaching a success state.  \emph{Reward testing}
was introduced in \cite{JHW94}; here the success states are labelled
by nonnegative real numbers---\emph{rewards}---to indicate degrees of
success, and reaching a success state
accumulates the associated reward. In \cite{Seg96} an infinite set of
success actions is used to report success, and the testing
outcomes are vectors of probabilities of performing these success
actions.  Compared to \cite{JHW94} this amounts to distinguishing
different qualities of success, rather than different quantities.

In \cite{WL92} and \cite{Seg96}, both tests and testees are
nondeterministic probabilistic processes, whereas \cite{JHW94} allows
nonprobabilistic tests only, thereby obtaining a less discriminating
form of testing. In \cite{esop} we strengthened reward testing by also
allowing probabilistic tests. Taking rewards testing in this form we showed that for finitary
processes, i.e. finite-state and finitely branching processes, all
three modes of testing lead to the same testing preorders. Thus, vector-based testing is no more powerful than \emph{scalar} testing that employs
only one success action, and likewise reward testing is no more
powerful than the special case of reward testing in which all rewards are 1.
\footnote{In spite of this there \emph{is} a difference in power
between the notions of testing from \cite{WL92} and \cite{Seg96}, but
this is an issue that is entirely orthogonal to the distinction
between scalar testing, reward testing and vector-based testing. In
\cite{Seg96} it is the execution of a success \emph{action} that
constitutes success, whereas in \cite{DNH84,henn,WL92,JHW94} it is
reaching a success \emph{state} (even though typically success actions
are used to identify those states). In \cite[Ex 5.3]{lmcs} we showed
that state-based testing is (slightly) more powerful than action-based
testing. The results presented in \cite{esop} about the coincidence of
scalar, reward, and vector-based testing preorders pertain to
action-based version of each, but in the conclusion it is observed
that the same coincidence could be obtained for their state-based
versions.  In the current paper we stick to state-based testing.}

In certain situations it is natural to introduce negative rewards.
This is the case, for instance, in the theory of Markov Decision
Processes \cite{Put94}. Intuitively, we could understand negative
rewards as costs, while positive rewards are often viewed as benefits
or profits.  This leads to the question: \emph{if negative rewards are
also allowed, how would the original reward-testing semantics change?}
We refer to the more relaxed form of testing, using positive and
negative rewards, as \emph{real-reward testing} and the original one
(from \cite{JHW94}, but with probabilistic tests as in \cite{esop}) as
\emph{nonnegative-reward testing}.

\begin{figure}
\begin{center} \tiny
\includegraphics[width=11cm,height=5cm]{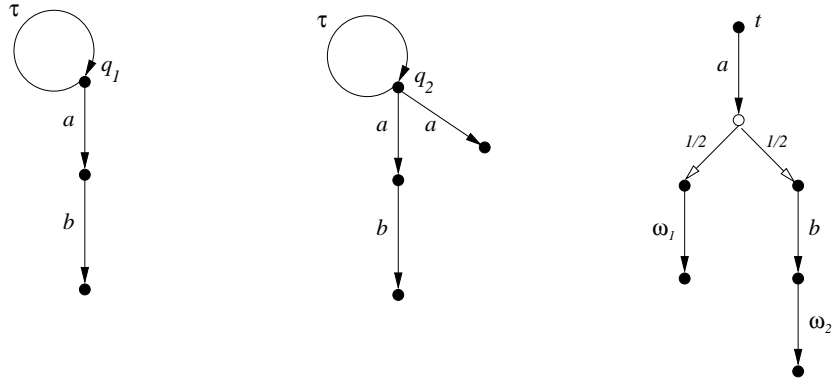}
\end{center}
\vspace{-1em}
\caption{Two processes with divergence and a test} \label{fig:diverg}
\end{figure}

The power of real-reward testing is illustrated in Figure~\ref{fig:diverg}.
The two (nonprobabilistic) processes in the left- and central diagrams are equivalent under
(probabilistic) may- as well as must testing; the $\tau$-loops in the
initial states cause both processes to fail any nontrivial must test. Yet, if a
reward of $-1$ is associated with performing the action $a$, and a
reward of $2$ with the subsequent performance of $b$
(implemented by the test in the right diagram; see Example~\ref{ex:divergence} for more details), in the first process the net
reward is either $0$ (if the process remains stuck in its initial
state) or positive, whereas running the second process may yield a
loss. This example shows that for processes that may exhibit
divergence, real-reward testing is more discriminating than
nonnegative-reward testing, or other forms of probabilistic testing.
It also illustrates that the extra power may be relevant in applications.

As remarked, in \cite{esop} we established that for finitary processes the
nonnegative-reward must-testing preorder ($\nrMustleq$) coincides
with the probabilistic must-testing preorder ($\pMustleq$), and
likewise for the may preorders.
Here we show that, in
contrast to the situation for nonnegative-reward (or scalar)
testing, for real-reward testing the may- and must preorders are the
inverse of each other, i.e. for any processes $\Delta$ and $\Gamma$,
\begin{equation}\label{eq:rrmay.rrmust}
\Delta \rrMayleq \Gamma \WIDERM{iff} \Gamma \rrMustleq \Delta.
\end{equation}
Our main result is that restricted to finitary convergent
processes, the real-reward must preorder coincides with the
nonnegative-reward must preorder, i.e. for any finitary convergent
processes $\Delta$, $\Gamma$,
\begin{equation}\label{eq:rrmust.nrmust}
\Delta \rrMustleq \Gamma \WIDERM{iff} \Delta\nrMustleq \Gamma.
\end{equation}
Here by convergence we mean that
there is no infinite sequence of internal transitions
of the form $\Lambda_0\ar{\tau}\Lambda_1\ar{\tau}\cdots$ with
distribution $\Lambda_0$ (and thus its successors) reachable from either $\Delta$ or $\Gamma$.
This rules out the processes of Figure~\ref{fig:diverg}.
Although it is easy to see that in (\ref{eq:rrmust.nrmust}) the former
implies the latter, to prove the opposite is far from trivial.
We employ a novel characterisation of
the usual resolution-based testing approach, without introducing concepts
like \emph{policy} \cite{Put94}, \emph{adversary} \cite{KNP04},
\emph{scheduler} \cite{Seg95} or \emph{resolution} \cite{esop} that are
external to the process under investigation; instead we describe the
mechanism for gathering test results in terms of the \emph{weak
$\tau$-moves} or \emph{derivations} \cite{concur} the investigated
process can make, and hence speak of \emph{derivation-based testing}.

This allows us to exploit the failure simulation preorder
$\failsimpreo$ that in \cite{concur} was proven to coincide with the
probabilistic must testing preorder $\pMustleq$ based on resolutions,
at least for finitary processes.
Using the derivational characterisation we can show that, for
finitary convergent processes, $\failsimpreo$ is contained in $\rrMustleq$.
Convergence is essential here, even though it is not needed to
establish that $\failsimpreo$ is contained in $\nrMustleq$.
Combining this with the results from \cite{esop} and \cite{concur} mentioned above
leads to our required result that $\nrMustleq$ is included in
$\rrMustleq$, as far as finitary convergent processes are
concerned. Consequently, in this case, all the relations of
Figure~\ref{fig:preorders} collapse into one.

\begin{figure}[t]
\begin{center}
\vspace{-1em}
$$\begin{array}{l@{\quad}l@{\quad}l@{\quad}l@{\quad}c@{\quad}l@{\quad}l@{\quad}l@{\quad}l}
(\rrMayleq)^{-1} & \Equ{\rm \Thm{rrmay.rrmust}} & \rrMustleq &
  \Equ{\rm\Thm{rrm.nrm}} &\nrMustleq & \Equ{\cite{esop}} & \pMustleq &
  \Equ{\cite{concur}} & \failsimpreo
\end{array}$$
\textit{\small The symbol $=$ between two relations means that they
  coincide for finitary convergent processes.}
\end{center}
\caption{The relationship of different testing preorders.}\label{fig:preorders}
\end{figure}

The rest of this paper is organised as follows. We start by recalling
notation for probabilistic labelled transition systems.  In
Section~\ref{sec:testing} we review the resolution-based testing
approach and show that the real-reward may preorder is simply the
inverse of the real-reward must preorder. Moreover, using the example
of Figure~\ref{fig:diverg}, we show that in the presence of divergence
the inclusion of $\nrMustleq$ in $\rrMustleq$ is proper. In
Section~\ref{sec:derivation} we present the derivation-based testing approach
and also show that the two approaches agree.
Then in Section~\ref{sec:agree} we show
for finitary convergent processes that real-reward must testing
coincides with nonnegative-reward must testing.
We conclude in Section~\ref{sec:conclusion}.

Due to lack of space, we omit all proofs: they are reported in
\cite{full}. Besides the related work already mentioned above, many
other studies on probabilistic testing and simulation semantics
have appeared in the literature. They are reviewed in
\cite{gdpf,lmcs}.

\section{Probabilistic Processes}\label{sec:lang}

A (discrete) probability \emph{subdistribution} over a set $S$ is a
function $\Delta: S \rightarrow [0,1]$ with $\sum_{s\inp S}\Delta(s)
\leq 1$; the \emph{support} of such a $\Delta$ is
$\support{\Delta}\Defs\setof{s \inp S}{\Delta(s) > 0}$, and its
\emph{mass} $|\Delta|$ is
\plat{$\sum_{s\inp\support{\Delta}}\Delta(s)$}.  A subdistribution is
a (total, or full) \emph{distribution} if $|\Delta| = 1$.  The point
distribution $\pdist{s}$ assigns probability $1$ to $s$ and $0$ to all
other elements of $S$, so that $\support{\pdist{s}}=\sset{s}$.  With
$\subdist{S}$ we denote the set of subdistributions over $S$, and with
$\dist{S}$ its subset of full distributions.

Let $\sset{\Delta_k \mid k \in K}$ be a set
of subdistributions, possibly infinite. Then $\sum_{k \in K} \Delta_k$ is the
real-valued function in $S \rightarrow \Real$ defined by
$(\sum_{k \in K} \Delta_k)(s) \mathrel{:=} \sum_{k \in K} \Delta_k(s)$.
This is a partial operation on subdistributions because
for some state $s$ the sum of $\Delta_k(s)$ might exceed $1$. If the
index set is finite, say $\sset{1..n}$, we often write $\Delta_1 +
\ldots + \Delta_n$.  For $p$ a real
number from $[0,1]$ we use
\begin{math}
  p\Times \Delta
\end{math}
to denote the subdistribution given by $(p\Times \Delta)(s)
\mathrel{:=} p\Times\Delta(s)$. Finally we use $\eDis$ to denote the
everywhere-zero subdistribution that thus has empty support. These
operations on subdistributions do not readily adapt themselves to
distributions; yet if $\sum_{k \in K} p_k \mathbin= 1$ for
some $p_k \geq 0$, and the $\Delta_k$ are distributions,
then so is $\sum_{k \in K}p_k \Times \Delta_k$.

The expected value $\sum_{s\in S} \Delta(s) \Times f(s)$ over a
subdistribution $\Delta$ of a bounded nonnegative function $f$ to the
reals or tuples of them is written $\ExpDeltaf$, and the image of a
subdistribution $\Delta$ through a function $f:S\rightarrow T$, for some set $T$, is written
$\Imgf{\Delta}$ --- the latter is the subdistribution over
$T$ given by $\Imgf{\Delta}(t):=\sum_{f(s)=t}\Delta(s)$ for each $t\in T$.
\newpage

\begin{definition} \rm
A \emph{probabilistic labelled transition system} (pLTS) is a triple
$\langle S, \Act, \rightarrow \rangle$, where\vspace{-3pt}
\begin{enumerate}[(i)] \parskip 0pt \itemsep 0pt
\item $S$ is a set of states,
\item $\Act$ is a set of visible actions,
\item relation $\rightarrow$ is a subset of $S \times \Act_\tau \times \dist{S}$.
\end{enumerate}
Here $\Act_\tau$ denotes $\Act\cup \{\tau\}$, where
$\tau \not\in\Act$ is the invisible- or internal action.
\end{definition}
\noindent
A (nonprobabilistic) labelled transition system (LTS) may be viewed
as a degenerate pLTS --- one in which only point distributions are
used.
In this paper a \emph{(probabilistic) process} will simply be a
distribution over the state set of a pLTS\@.
As with LTSs, we write $s \ar{\alpha} \Delta$ for
$(s,\alpha,\Delta) \inp \mathord\rightarrow$, as well as
$s\ar{\alpha}$ for $\exists\Delta: s\ar{\alpha}\Delta$ and
$s\!\rightarrow$ for $\exists\alpha\!: s\,\ar{\alpha}$,
with $s\nar{\alpha}$ and $s \not\!\rightarrow$ representing their negations.
A pLTS is \emph{deterministic} if for any state $s$ and
label $\alpha$ there is at most one distribution $\Delta$ with $s\ar{\alpha}\Delta$.
It is \emph{finitely branching} if the set $\sset{\Delta\mid
s\ar{\alpha}\Delta,~ \alpha\inp L}$ is finite for all states $s$;
if moreover $S$ is finite, then the pLTS is \emph{finitary}.
A subdistribution $\Delta$ over the state set $S$ of an arbitrary pLTS is \emph{finitary}
if restricting $S$ to the states reachable from $\Delta$
yields a finitary sub-pLTS.

\section{Testing probabilistic processes}\label{sec:testing}

A \emph{test} is a distribution over the state set of a pLTS having 
$\Act_\tau \cup \Omega$ as its set of transition labels, where $\Omega$ is a
set of fresh \emph{success} actions, not already in $\Act_\tau$, introduced specifically to
report testing outcomes.\footnote{For \emph{vector-based} testing we
  normally take $\Omega$ to be countably infinite \cite{Seg96}.  This
  way we have an unbounded supply of success actions for
  building tests, of course without obligation to use them all.
  \emph{Scalar} testing is obtained by taking $|\Omega|=1$.}
For simplicity we may assume a fixed pLTS of
processes---our results apply to any choice of such a pLTS---and a
fixed pLTS of tests.  Since the power of testing depends on the expressivity
of the pLTS of tests---in particular certain types of tests are
necessary for our results---let us just postulate that this pLTS is sufficiently expressive
for our purposes --- for example that it can be used to interpret all
processes from the language \pCSP, as in our previous papers
\cite{gdpf,lmcs,concur}.

Although we use success \emph{actions}, they are used merely to mark
certain states as success states, namely the sources of transitions labelled by success
actions. For this reason we systematically ignore
the distributions that can be reached after a success action.
We impose two requirements on all states in a pLTS of tests, namely
\begin{itemise}
\item[(A)]
if $t \ar{\omega_1}$ and $t \ar{\omega_2}$ with $\omega_1,\omega_2\in\Omega$
then $\omega_1 = \omega_2$.
\item[(B)]
if $t \ar{\omega}$ with $\omega\in\Omega$ and $t \ar{\alpha} \Delta$ with
$\alpha\in\Act_\tau$ then $u\ar{\omega}$ for all $u\in\support{\Delta}$.
\end{itemise}
The first condition says that a success state can have one success
identity only, whereas the second condition is slight weakening
of the requirement from \cite{JHW94} that success states must be end
states; it allows further progress from an $\omega$-success state, for
some $\omega\in\Omega$, but $\omega$ must remain enabled.
\footnote{Justification for imposing such restrictions can be found in Appendix A of \cite{esop}.}

To apply test $\Theta$ to process $\Delta$ we form a parallel
composition $\Theta \parT \Delta$ in which \emph{all} visible actions
of $\Delta$ must synchronise with $\Theta$. The synchronisations are
immediately renamed into $\tau$.  The resulting composition is a
process whose only possible actions are the elements of $\Omega_\tau
:=\Omega \cup \{\tau\}$.  Formally, if $\langle {\bf P}, \Act,
\rightarrow_{\bf P} \rangle$ and $\langle {\bf T}, \Act \cup
\Omega, \rightarrow_{\bf T} \rangle$ are the pLTSs of processes and
tests, then the pLTS of applications of tests to processes is $\langle
{\bf C}, \Omega, \rightarrow \rangle$, with ${\bf C} =
\{t\|p \mid t \inp {\bf T} \wedge p\inp {\bf P}\}$ and $\rightarrow$ the
transition relation generated by the rules in Fig.~\ref{fig:par}.
Here if $\Theta\in\dist{\bf T}$ and $\Delta\in\dist{\bf P}$,
then $\Theta \| \Delta$ is the distribution given by $(\Theta \|
\Delta)(t\|p)\mathrel{:=} \Theta(t)\cdot\Delta(p)$.
The resulting pLTS also satisfies (A), (B) above;
this would not be the case if we had strengthened
(B) to require that success states must be end states.

\begin{figure}[t]
$$
   \linferSIDE[\Rlts{par.l}]{t \ar{\alpha}_{\bf T} \Theta}
                 {t \Par{A} p \ar{\alpha} \Theta \Par{A} \pdist{p}}
                 {~~~~~\alpha \mathop{\not\in} \Act}
\qquad
   \linferSIDE[\Rlts{par.r}]{p \ar{\alpha}_{\bf P} \Delta}
                 {t \Par{A} p \ar{\alpha} \pdist{t} \Par{A} \Delta}
                 {~~~~~\alpha \mathop{\not\in} \Act}
\qquad
   \makebox[4.5cm][l]{$\linferSIDE[\Rlts{par.i}]{t \ar{a}_{\bf T} \Theta\;\;\;p \ar{a}_{\bf P} \Delta\;\;}
                {t \Par{A} p\ar{\tau} \Theta \Par{A} \Delta}
        {a\inp \Act}$}
$$
\caption{Synchronous parallel composition between tests and processes\label{fig:par}}
\end{figure}

We will define the result $\RApply(\Theta,\Delta)$ of applying the test
$\Theta$ to the process $\Delta$ to be a set of testing outcomes, exactly
one of which results from each resolution of the choices in $\Theta \parT \Delta$.
\pagebreak[3]
Each \emph{testing outcome} is
an $\Omega$-tuple of real numbers in the interval
[0,1], i.e.\ a function $o:\Omega\rightarrow [0,1]$, and its
$\omega$-component $o(\omega)$, for $\omega \in \Omega$, gives the
probability that the resolution in question will reach an
\emph{$\omega$-success state}, one in which the success action
$\omega$ is possible.

Due to the presence of nondeterminism in pLTSs, we need a mechanism to
reduce a nondeterministic structure into a set of deterministic
structures, each of which determines a single possible outcome. Here we adapt
the notion of \emph{resolution}, defined in \cite{esop} for
probabilistic automata, to pLTSs.

\begin{definition} \bf[Resolution]\rm\label{def:resolution}
A \emph{resolution} of a subdistribution $\Delta \inp\subdist{S}$ in a
pLTS  $\langle S, {\Omega}, \rightarrow \rangle$  is a triple
 $\langle R, \Lambda, \rightarrow_R \rangle$
where
$\langle R, {\Omega}, \rightarrow_R \rangle$ is a deterministic pLTS and
$\Lambda \inp \subdist{R}$, such that there exists a
\emph{resolving function} $f: R\rightarrow S$ satisfying
\begin{enumerate}[(i)]
\item $\Imgf{\Lambda} = \Delta$
\item if $r\ar{\alpha}_R\Lambda'$ for $\alpha\in {\Omega}_\tau$
      then $f(r)\ar{\alpha}\Imgf{\Lambda'}$
\item if $f(r)\ar{\alpha}$ for $\alpha\in {\Omega}_\tau$ then $r\ar{\alpha}_R$\;.
 \end{enumerate}
\end{definition}
\noindent
The reader is referred to Section 2 of \cite{esop} for a detailed
discussion of the concept of resolution, and the manner in which a
resolution represents a run of a process; in particular in a
resolution states in $S$ are allowed to be resolved into distributions,
and computation steps can be \emph{probabilistically interpolated}.
Our resolutions match the results of applying a scheduler as defined
in \cite{Seg95}.

We now explain how to associate  an outcome with a particular
resolution, which in turn will associate a set of outcomes with a
subdistribution in a pLTS\@. Given a deterministic pLTS $\langle R,
{\Omega}, \rightarrow_R \rangle$ consider the functional
\begin{math}\label{page:resolutions}
  {\cal R}: (R \rightarrow [0,1]^\Omega) \rightarrow  (R \rightarrow [0,1]^\Omega)
\end{math}
defined by
\begin{align}\label{eq:functional}
   {\cal R}(f)(r)(\omega) \mathrel{:=}
\begin{cases}
1 &\mbox{if $r \ar{\omega}$}\\
0 &\mbox{if $r \nar{\omega}$ and $r \nar{\tau}$}\\
\ExpDeltaf(\omega) &\mbox{if $r \nar{\omega}$ and $r \ar{\tau}
\Delta$.}
\end{cases}
\end{align}
We view the unit interval $[0,1]$ ordered in
the standard manner as a complete lattice;
this induces the structure of a complete lattice on the product   $ [0,1]^\Omega$ and in turn
on the  set of functions
$R \rightarrow [0,1]^\Omega$.  The functional ${\cal R}$ is easily seen to be
monotonic and therefore has a least fixed point, which we denote by
$\ResultsN_{\langle R, {\Omega}, \rightarrow_R \rangle}$; this is abbreviated to $\ResultsN$ when the
deterministic pLTS in question is understood.

Now we define $\RApply(\Theta,\Delta)$ to be the set of vectors
\begin{equation}\label{eq:apply}
\RApply(\Theta,\Delta)~:=~\setof{\Exp{\Lambda}(\ResultsN_{\langle R, {\Omega}, \rightarrow_R \rangle})}
{\langle R, \Lambda, \rightarrow_R \rangle \mbox{ is a resolution of }  \Theta \| \Delta
}\;.
\end{equation}

\begin{figure}[t]
\begin{center} \tiny
\includegraphics[width=8cm,height=5.5cm]{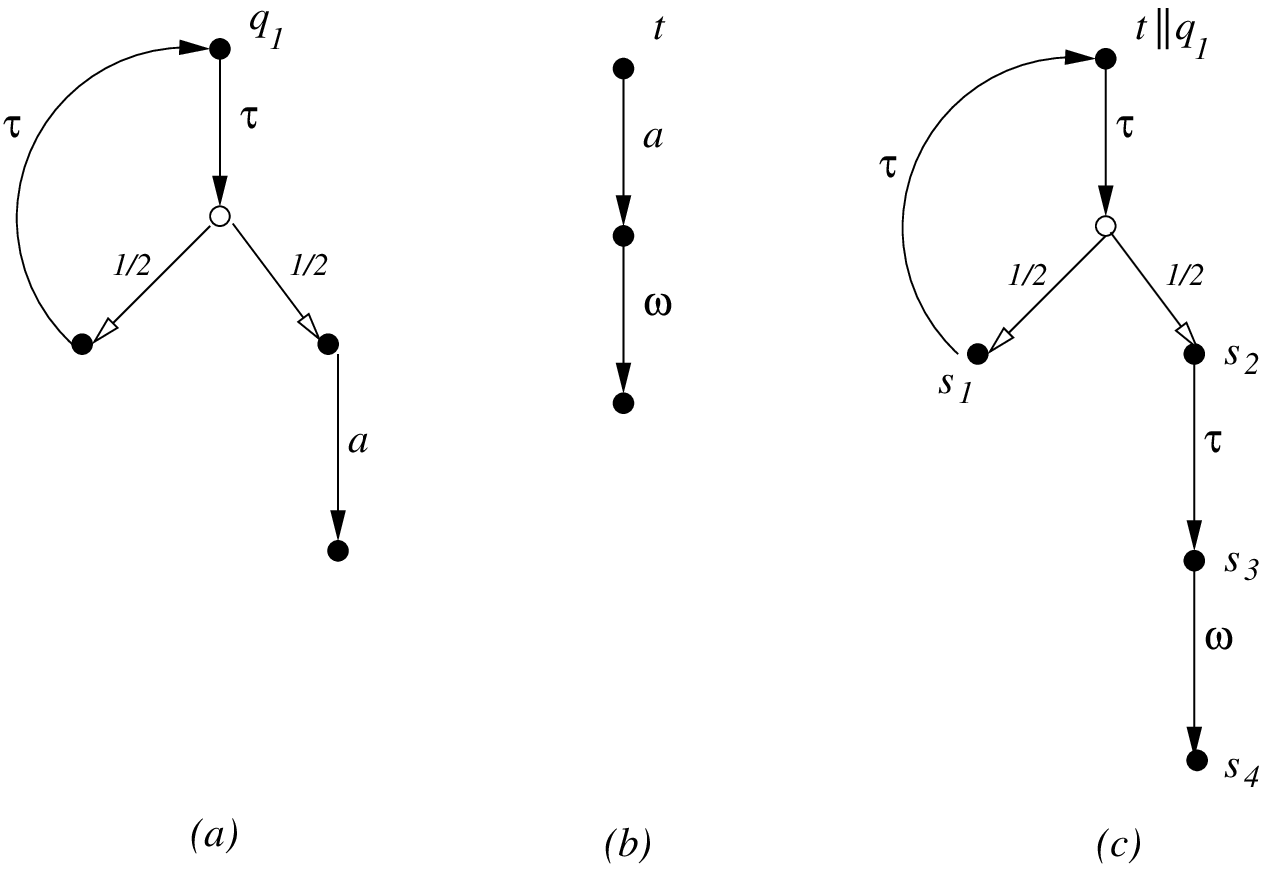}
\end{center}
\caption{Testing the  process $\pdist{q_1}$} \label{fig:ex1B}
\end{figure}

\begin{example}\rm\label{ex:ex1B}
  Consider the process $\pdist{q_1}$
  depicted in Figure~\ref{fig:ex1B}(a). Here states are
represented by filled nodes $\bullet$ and distributions by open nodes $\circ$.
  We leave out point-distributions --- diverting an incoming
edge to the unique state in its support.
When we apply the test
  $\pdist{t}$ depicted in Figure~\ref{fig:ex1B}(b) to it we get the process
  $\pdist{t\|q_1}$ depicted in  Figure~\ref{fig:ex1B}(c).
This process is already deterministic, hence has essentially only one resolution: itself.
Moreover the outcome {$\Exp{\pdist{t\|q_1}}(\ResultsN) =
  \ResultsN(t\|q_1)$} associated with it is
the least solution of the equation
\begin{math}
  \ResultsN(t\|q_1) = \frac{1}{2}\cdot \ResultsN(t\|q_1) + \frac{1}{2}\vec{\omega}
\end{math}
where $\vec{\omega}: \Omega\rightarrow[0,1]$ is the $\Omega$-tuple
with $\vec{\omega}(\omega)=1$ and $\vec{\omega}(\omega')=0$ for all
$\omega'\not=\omega$.
In fact this equation has a unique solution in $[0,1]^\Omega$, namely $\vec{\omega}$.
Thus $\RApply(\pdist{t},\pdist{q_1}) = \sset{\vec{\omega}}$.  
\end{example}

\begin{example}\rm\label{ex:ex2B}
  Consider the process
$\pdist{q_2}$ and the application of the test $\pdist{t}$ to it, as
outlined in Figure~\ref{fig:ex2}.
For each $k \geq 1$ the process $\pdist{t \parT q_2}$ has a resolution
$\langle R_k, \Lambda, \rightarrow_{R_k} \rangle$ such that
$\Exp{\Lambda}(\ResultsN) = (1 \mathord-\frac{1}{2^k})\vec{\omega}$; intuitively it goes around
the loop $(k-1)$ times before at last taking the right hand $\tau$
action.  Thus $\RApply(\pdist{t},\pdist{q_2})$ contains $(1 - \frac{1}{2^k})\vec{\omega}$ for
every $k \geq 1$. But it also contains $\vec{\omega}$, because of the
resolution which takes the left hand $\tau$-move every time. Thus
$\RApply(\pdist{t},\pdist{q_2})$ includes the set
\begin{align*}\textstyle
\sset{(1 \mathord-\frac{1}{2})\vec{\omega},\;
      (1\mathord-\frac{1}{2^2})\vec{\omega},\ldots, (1 \mathord-\frac{1}{2^k})\vec{\omega}, \ldots, \vec{\omega}}
\end{align*}
As resolutions allow any interpolation between the two $\tau$-transitions
from state $s_1$,
$\RApply(\pdist{t},\pdist{q_2})$ is actually the convex closure of the above set.
\begin{figure}[t]
\begin{center}
\includegraphics[width=12cm,height=6cm]{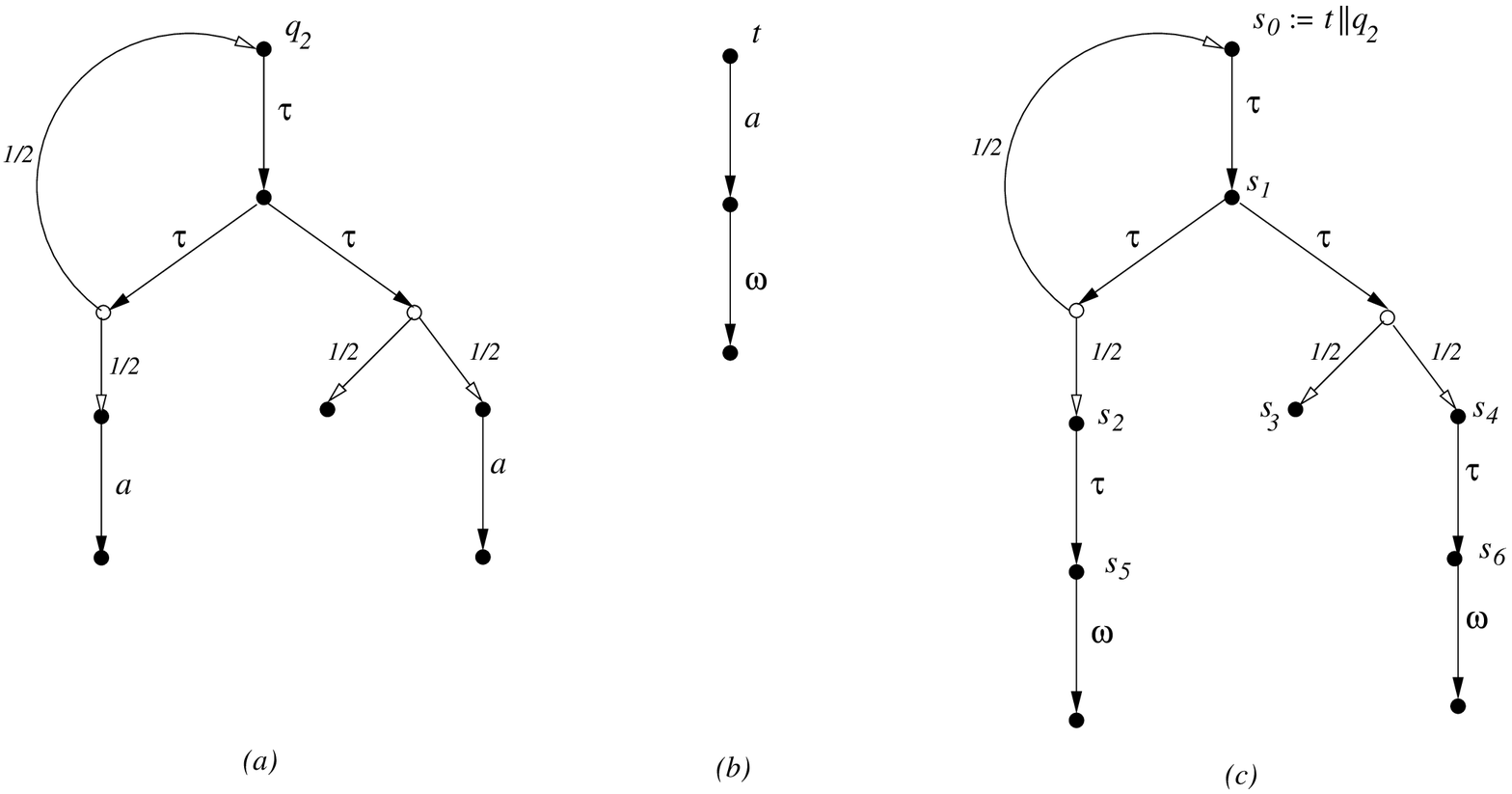}
\end{center}
\caption{Testing the process $\pdist{q_2}$} \label{fig:ex2}
\end{figure}
\end{example}

\noindent
There are two standard methods for comparing two sets of ordered outcomes:
\vspace{-5pt}
\begin{align*}
  O_1 \Hleq O_2  &\qquad\text{if for every $o_1 \in O_1$ there exists some $o_2 \in O_2$ such that $o_1 \leq o_2$}
\\[-3pt]
  O_1 \Sleq O_2  &\qquad\text{if for every $o_2 \in O_2$ there exists some $o_1 \in O_1$ such that $o_1 \leq o_2$}
\end{align*}
This gives us our definition of the probabilistic may- and must-testing preorders;
they are decorated with $\Times^\Omega$ for
the repertoire $\Omega$ of testing actions they employ.

\begin{definition} \bf[Probabilistic testing preorders]\rm\label{def:d1445}\qquad
  \begin{enumerate}[(i)]
  \item $\Delta \pmMayleq \Gamma$ if for every $\Omega$-test $\Theta$, $\RApply(\Theta,\Delta) \Hleq \RApply(\Theta,\Gamma) $.
  \item $\Delta \pmMustleq \Gamma$ if for every $\Omega$-test $\Theta$, $\RApply(\Theta,\Delta) \Sleq \RApply(\Theta,\Gamma) $.
\end{enumerate}
These preorders are abbreviated to $\Delta \pMayleq \Gamma$ and $\Delta \pMustleq \Gamma$ when $\size{\Omega} = 1$.
\end{definition}

In \cite{esop} we established that for finitary processes $\pmMayleq$
coincides with $\pMayleq$ and $\pmMustleq$ with $\pMustleq$ for any
choice of $\Omega$.  We also defined the reward testing preorders in
terms of the mechanism set up so far. The idea is to associate with
each success action $\omega\in \Omega$ a reward, which is a nonnegative
number in the unit interval $[0,1]$; and then a run of a probabilistic
process in parallel with a test yields an expected reward accumulated
by those states which can enable success actions. A reward tuple $h\in
[0,1]^\Omega$ is used to assign reward $h(\omega)$ to success action
$\omega$, for each $\omega\in\Omega$.
Due to the presence of nondeterminism, the application of a test
$\Theta$ to a process $\Delta$ produces a set of expected rewards.
Two sets of rewards can be compared by examining their suprema/infima; this gives
us two methods of testing called reward may/must testing. In
\cite{esop} all rewards are required to be nonnegative, so we refer to
that approach of testing as \emph{nonnegative-reward testing}. If we
also allow negative rewards, which intuitively can be understood as
costs, then we obtain an approach of testing called \emph{real-reward
  testing}. Technically, we simply let reward tuples $h$ range over
the set $[-1,1]^\Omega$. If $o\in [0,1]^\Omega$, we use the
dot-product $h\cdot o=\sum_{\omega\in\Omega} h(\omega)\cdot
o(\omega)$. It can apply to a set $O\subseteq [0,1]^\Omega$ so that
$h\cdot O=\sset{h\cdot o\mid o\in O}$. Let $A\subseteq [-1,1]$. We use
the notation $\bigsqcup A$ for the supremum of set $A$, and $\bigsqcap
A$ for the infimum.
\begin{definition} \bf[Reward testing preorders]\rm\label{def:d1509}\qquad
 \begin{enumerate}[(i)]
  \item $\Delta \nrmMayleq \Gamma$ if for every $\Omega$-test $\Theta$ and nonnegative-reward tuple $h\in [0,1]^\Omega$,\\ $\bigsqcup h\cdot\RApply(\Theta,\Delta) \leq \bigsqcup h\cdot \RApply(\Theta,\Gamma) $.
  \item $\Delta \nrmMustleq \Gamma$ if for every $\Omega$-test $\Theta$ and nonnegative-reward tuple $h\in [0,1]^\Omega$,\\ $\bigsqcap h\cdot\RApply(\Theta,\Delta) \leq \bigsqcap h\cdot\RApply(\Theta,\Gamma) $.
  \item $\Delta \rrmMayleq \Gamma$ if for every $\Omega$-test $\Theta$ and real-reward tuple $h\in [-1,1]^\Omega$,\\ $\bigsqcup h\cdot\RApply(\Theta,\Delta) \leq \bigsqcup h\cdot\RApply(\Theta,\Gamma) $.
  \item $\Delta \rrmMustleq \Gamma$ if for every $\Omega$-test $\Theta$ and real-reward tuple $h\in [-1,1]^\Omega$,\\ $\bigsqcap h\cdot\RApply(\Theta,\Delta) \leq \bigsqcap h\cdot\RApply(\Theta,\Gamma) $.
\end{enumerate}
This time we drop the superscript $\Omega$ iff $\Omega$ is countably infinite.
\end{definition}

\noindent
It is shown in Corollary 1 of \cite{esop} that nonnegative-reward testing is equally powerful as probabilistic testing.
\begin{theorem} {\bf\cite{esop}}\label{thm:nr.pb}
For any finitary processes $\Delta$ and $\Gamma$,
\begin{enumerate}[(i)]
\item $\Delta \nrMayleq \Gamma$ if and only if $\Delta \pMayleq \Gamma$.
\item $\Delta \nrMustleq \Gamma$ if and only if $\Delta \pMustleq \Gamma$.
\end{enumerate}
\end{theorem}

\noindent
In this paper we focus on the real-reward testing preorders
$\rrMayleq$ and $\rrMustleq$, by comparing them with the nonnegative
reward testing preorders $\nrMayleq$ and $\nrMustleq$.
Although these two nonnegative-reward testing preorders are in
general incomparable we have:

\begin{theorem}\label{thm:rrmay.rrmust}
For any processes $\Delta$ and $\Gamma$, it holds that
 $\Delta \rrMayleq \Gamma$ if and only if $\Gamma \rrMustleq \Delta$.
\end{theorem}

\noindent
Our next task is to compare $\rrMustleq$ with $\nrMustleq$. The former
is included in the latter, which directly follows from
Definition~\ref{def:d1509}. Surprisingly, it turns out that for
finitary convergent processes the latter is also included in the
former, thus establishing that the two preorders are in fact the same. The rest of the
paper is devoted to proving this result. However, we first show that
this result does not extend to divergent processes.

\begin{example}\label{ex:divergence}
Consider the processes $\pdist{q_1}$ and $\pdist{q_2}$ depicted in Figure~\ref{fig:diverg}.
Using the characterisations of $\pMayleq$ and $\pMustleq$ in \cite{concur},
it is easy to see that these processes cannot be distinguished by
probabilistic may- and must testing, and hence not by nonnegative-reward testing either.
However, let $\pdist{t}$ be the test in the right diagram of Figure~\ref{fig:diverg} that first synchronises on the action
$a$, and then with probability $\frac{1}{2}$ reaches a state in which a reward
of $-2$ is allocated, and with the remaining probability $\frac{1}{2}$
synchronises with the action $b$ and reaches a state that yields a reward of $4$.
Thus the test employs two success actions $\omega_1$ and $\omega_2$, and we use
the reward tuple $h$ with $h(\omega_1)=-2$ and $h(\omega_2)=4$.
Then the resolution of $\pdist{q_1}$ that does not involve the
$\tau$-loop contributes the value  $-2\cdot \frac{1}{2} + 4\cdot \frac{1}{2} = -1+2 = 1$
to the set $h\cdot\RApply(\pdist{t},\pdist{q_1})$, whereas the
resolution that only involves the $\tau$-loop contributes the value $0$.
Due to interpolation, $h\cdot\RApply(\pdist{t},\pdist{q_1})$ is in fact
the entire interval $[0,1]$. On the other hand, the resolution
corresponding to the $a$-branch of $q_2$ contributes the value $-1$ and
$h\cdot\RApply(\pdist{t},\pdist{q_2}) = [-1,1]$.
Thus 
$\bigsqcap h\cdot\RApply(\pdist{t},\pdist{q_1}) = 0 >
 -1 = \bigsqcap h\cdot\RApply(\pdist{t},\pdist{q_2})$,
and hence $\pdist{q_1} \not \rrMustleq \pdist{q_2}$.
\end{example}

\section{Derivation-based testing}\label{sec:derivation}
In this section we give an alternative definition of $\RApply(\Theta,\Delta)$.
Our definition has four ingredients. First of all, for technical reasons we normalise our
pLTS of applications of tests to processes by
pruning away all outgoing $\tau$-transitions from success states. This
way an $\omega$-success state will only have outgoing transitions
labelled $\omega$.

\begin{definition} \bf[$\omega$-respecting] \rm A pLTS $\langle S, \Omega, \rightarrow \rangle$
is said to be \emph{$\omega$-respecting} whenever $s \ar{\omega}$, for
any $\omega \in \Omega$, implies $s \nar{\tau}$.
\end{definition}
\noindent
It is straightforward to modify the pLTS of applications of tests to
processes into one that it is $\omega$-respecting, namely by removing
all transitions $s \ar{\tau}\Delta$ for states $s$ with $s \ar{\omega}$.
With $[\Theta\parT\Delta]$ we denote the distribution
$\Theta\parT\Delta$ in this pruned pLTS.

Secondly, we recall the definition of weak derivations from \cite{concur}.
In a pLTS actions are only performed by states, in that actions are
given by relations from states to distributions. But processes
in general correspond to distributions over states, so in order to
define what it means for a process to perform an action, we need to
\emph{lift} these relations so that they also apply to
distributions. In fact we will find it convenient to lift them to
subdistributions.
\begin{definition}\rm\label{def:lift}
Let $(S,L,\rightarrow)$ be a pLTS and $\mathord{\aRel} \subseteq
  S\times\subdist{S}$ be a relation from states to subdistributions.
Then $\mathord{\lift{\aRel}} \subseteq \subdist{S} \times \subdist{S}$
is the smallest relation that satisfies:
\begin{enumerate}[(i)]
\item $s \aRel \Delta$ implies $\pdist{s} \lift{\aRel} \Delta$, and
\item\label{i1529} (Linearity)
 $\Gamma_i \lift{\aRel} \Delta_i$ for $i\inp I$ implies
 $(\sum_{i\in I}p_i\Times\Gamma_i)~\lift{\aRel}~(\sum_{i\in I}p_i\Times\Delta_i)$
 for any $p_i \inp [0,1]$ ($i\inp I$) with $\sum_{i\in I}p_i \leq 1$.
\end{enumerate}
\end{definition}
\noindent
An application of this notion is when the relation is $\ar{\alpha}$
for $\alpha\in \Act_\tau$; in that case we also write $\ar{\alpha}$
for $\lift{\raisebox{0pt}[0.7em][0pt]{$\ar{\alpha}$}}$. Thus, as source of a relation $\ar{\alpha}$ we
now also allow distributions, and even subdistributions.
A subtlety of this approach is that for any action $\alpha$, we have
$\eDis \ar{\alpha} \eDis$
simply by taking $I=\emptyset$ or $\sum_{i\in I}p_i=0$ in
Definition~\ref{def:lift}. That turns out to make $\eDis$ especially useful for
modelling the ``chaotic'' aspects of divergence in \cite{concur}, in
particular that in the must-case a divergent process can simulate any
other.
\pagebreak

\begin{definition} \bf[Weak derivation]\rm
\label{def:d0839}
Suppose we have subdistributions $\Delta, \Delta_k^{\rightarrow},
\Delta_k^{\Stop}$, for $k \geq 0$, with the following properties:
\vspace{-1.5em}
\label{weaktau}
\begin{eqnarray*}
  \Delta &=& \Delta_0^{\rightarrow} + \Delta_0^\Stop\\
  \Delta_0^{\rightarrow}  &\ar{\tau}& \Delta_1^{\rightarrow} + \Delta_1^\Stop\\[-7pt]
&\vdots \\
  \Delta_k^{\rightarrow}  &\ar{\tau}& \Delta_{k{+}1}^{\rightarrow} + \Delta_{k{+}1}^\Stop~.\\[-7pt]
&\vdots
\end{eqnarray*}
\vspace{-2em}

\noindent
Then we call $\Delta'\Defs \sum_{k=0}^\infty \Delta_k^{\Stop}$ a
\emph{weak derivative} of $\Delta$, and write $\Delta\dar{}\Delta'$
to mean that $\Delta$ can make a \emph{weak derivation} to its
derivative $\Delta'$.
\end{definition}
\noindent
There is always at least one weak derivative of any subdistribution (the
subdistribution itself) and there can be many.

Thirdly, we identify a class of special weak derivatives called extreme derivatives.
\begin{definition} \bf[Extreme derivatives]\rm\label{def:d1814}
A state $s$ in a pLTS is called \emph{stable} if $s\nar{\tau}$, and a
subdistribution $\Delta$ is called \emph{stable} if every state in its
support is stable.  We write $\Delta \darE{}\Delta'$ whenever $\Delta
\dar{} \Delta'$ and $\Delta'$ is stable, and call $\Delta'$ an
\emph{extreme} derivative of $\Delta$.
\end{definition}
\noindent
Referring to Definition~\ref{def:d0839}, we see this means that in the extreme
derivation of $\Delta'$ from $\Delta$ at every stage a state must move
on if it can, so that every stopping component can contain only states
which \emph{must} stop: for $s \in \support{\Delta_k^{\rightarrow} + \Delta_k^{\Stop}}$
we have $s \in \support{\Delta_k^{\Stop}}$ if \emph{and now also} only if $s\nar{\tau}$.
Moreover if the pLTS is $\omega$-respecting then whenever $s \in
\support{\Delta_k^\Move}$, it is not successful, i.e. $s \nar{\omega}$ for
every $\omega \in \Omega$.

\begin{lemma} {\bf[Existence of extreme derivatives]} \label{lem:unique.der}
  \begin{enumerate}[(i)]\item\label{item:i0834} For every
    subdistribution $\Delta$ there exists some (stable) $\Delta'$ such
    that $\Delta \darE{} \Delta'$.
\item\label{item:i1319} In a deterministic pLTS if $\Delta \darE{} \Delta'$ and $\Delta \darE{} \Delta''$ then $\Delta' = \Delta''$.
\end{enumerate}
\end{lemma}
\noindent
\emph{Sub}distributions are essential here.
Consider  a state $t$ that has only one transition, a self $\tau$-loop
$t\ar{\tau}\pdist{t}$.  Then it diverges and it has a unique
extreme derivative $\eDis$, the empty subdistribution. More generally, suppose a subdistribution
$\Delta$ diverges, that is there is an infinite sequence of
internal transitions \mbox{$\Delta \ar{\tau}
\Delta_1 \ar{\tau} \ldots \Delta_k \ar{\tau} \ldots$}.  Then one
extreme derivative of $\Delta$ is $\eDis$, but it may have others.

The final ingredient in the definition of a set of outcomes of an
application of a test to a process is the outcome of a particular
extreme derivative.
Note that all states $s \in \support{\Delta}$ in the support of an
extreme derivative either satisfy $s \ar{\omega}$ for a unique
$\omega\in\Omega$, or have $s \not\!\rightarrow$.

\begin{definition} {\bf[Outcomes]}\rm\label{def:immediate reward}
The outcome $\Rfun\Delta \in [0,1]^\Omega$ of a stable subdistribution
$\Delta$ is given by
$\Rfun\Delta(\omega) = \sum\sset{\Delta(s) \mid s \in \support{\Delta},~s \ar{\omega}}$.
\end{definition}
\noindent
Putting all four ingredients together, we arrive at a definition of
$\Apply(\Theta,\Delta)$:
\begin{definition}\rm\label{def:apply}
Let $\Delta$ be a process and $\Theta$ an $\Omega$-test.
Then $\Apply(\Theta,\Delta) = \sset{\Rfun\Lambda \mid [\Theta\parT \Delta]\darE{}\Lambda}.$
\end{definition}

\noindent
The role of pruning in the above definition can be seen via the
following example.
\begin{example} \rm
Let $\pdist{p}$ be a process that first does an $a$-action, to the
point distribution $\pdist{q}$, and then diverges, via the
$\tau$-loop $q \ar{\tau}\pdist{q}$. Let $\pdist{t}$ be the test used in
Examples~\ref{ex:ex1B} and~\ref{ex:ex2B}.  Then $\pdist{t}\parT
\pdist{p}$ has a unique extreme
derivative $\eDis$,
whereas $[\pdist{t}\parT \pdist{p}]$ has a unique extreme
derivative $[\omega\parT q]$.
Here we give the name $\omega$ to the state reachable from $\pdist{t}$
with the outgoing $\omega$-transition. The
outcome in $\Apply(\pdist{t},\pdist{p})$ shows that process
$\pdist{p}$ passes test $\pdist{t}$ with probability $1$, which is
what we expect for state-based testing.  Without pruning we would get
an outcome saying that $\pdist{p}$ passes $\pdist{t}$ with probability~$0$.
\end{example}
\noindent
As this example is nonprobabilistic, it also illustrates how pruning
enables the standard notion of nonprobabilistic testing to be
captured by derivation-based testing.

\begin{example}\rm\label{ex:ex1} (Revisiting Example~\ref{ex:ex1B}.)
The pLTS  in  Figure~\ref{fig:ex1B}(c) is deterministic and unaffected
by pruning; from part (\ref{item:i1319}) of Lemma~\ref{lem:unique.der}
it follows that $\pdist{t\parT q_1}$  has a unique extreme derivative
$\Lambda$. Moreover $\Lambda$ can be calculated to be
\plat{\begin{math}
  \sum_{k \geq 1} \frac{1}{2^k} \cdot \pdist{s_3},
\end{math}}
which simplifies to the distribution $\pdist{s_3}$. Therefore, $\Apply(\pdist{
t},\pdist{q_1})=\sset{\Rfun\pdist{s_3}}=\sset{\vec{\omega}}$.
\end{example}

\begin{example}\rm\label{ex:ex2} (Revisiting Example~\ref{ex:ex2B}.)
The application of the test $\pdist{t}$ to processes $\pdist{q_2}$
is outlined in Figure~\ref{fig:ex2}(c).
Consider any extreme derivative $\Delta'$ from $s_0 = [\pdist{t} \parT \pdist{q_2}]$;
note that here again pruning actually has no effect. Using the
notation of Definition~\ref{def:d0839}, it is clear that
$\Delta_0^\Stop$ and $\Delta_0^\Move$ must be $\eDis$ and $\pdist{s_0}$ respectively.
Similarly,  $\Delta_1^\Stop$ and $\Delta_1^\Move$ must be
$\eDis$ and $\pdist{s_1}$ respectively.
 But $s_1$ is a
nondeterministic state, having two possible transitions:
\begin{enumerate}[(i)]
\item $s_1 \ar{\tau} \Lambda_0$ where $\Lambda_0$ has support $\sset{s_0,s_2}$ and assigns each of them
the weight $\frac{1}{2}$

\item $s_1 \ar{\tau} \Lambda_1$ where $\Lambda_1$ has the support $\sset{s_3,s_4}$, again dividing the mass
equally among them.
\end{enumerate}
So  there are many possibilities for $\Delta_2$;
from Definition~\ref{def:d0839} one sees that in fact $\Delta_2$ can be of the form
\vspace{-1ex}
\begin{equation}\label{eq:choiceofp}
  p \cdot \Lambda_0 + (1-p) \cdot \Lambda_1
\vspace{-1ex}
\end{equation}
for any choice of $p \in [0,1]$.

Let us consider one possibility, an extreme one where $p$ is chosen to be $0$; only the transition
(ii) above is used.  Here $\Delta_2^\Move$ is the subdistribution $\frac{1}{2}\pdist{s_4}$, and
$\Delta_k^\Move = \eDis$ whenever $k >2$. A simple calculation shows that in this case the extreme
derivative generated is  $ \Lambda_1^e =
\frac{1}{2}\pdist{s_3}+\frac{1}{2}\pdist{s_6}$ which implies that
$\frac{1}{2}\vec{\omega} \in \Apply(\pdist{t},\pdist{q_2})$.

Another possibility for $\Delta_2$ is $\Lambda_0$, corresponding to
$p = 1$ in (\ref{eq:choiceofp}) above. Continuing
this derivation leads to $\Delta_3$ being $\frac{1}{2} \cdot
\pdist{s_1} + \frac{1}{2} \cdot \pdist{s_5}$; thus
$\Delta_3^\Stop = \frac{1}{2} \cdot  \pdist{s_5}$ and
$\Delta_3^\Move = \frac{1}{2} \cdot \pdist{s_1}$. Now in the
generation of $\Delta_4$ from $\Delta_3^\Move$ again we
resolve a transition from the nondeterministic state $s_1$, by
choosing some arbitrary $p \in [0,1]$ in (\ref{eq:choiceofp}).
Suppose we choose $p\mathbin=1$ every time, completely ignoring
transition (ii) above. Then the extreme derivative generated is
\vspace{-1ex}
\begin{equation*}
  \Lambda_0^e = \sum_{k \geq 1} \frac{1}{2^k} \cdot  \pdist{s_5}
\vspace{-1ex}
\end{equation*}
which simplifies to the distribution $\pdist{s_5}$. This in turn
means that $\vec{\omega} \in \Apply(\pdist{t},\pdist{q_2})$.

We have seen two possible derivations of extreme derivatives from
$\pdist{s_0}$.  But there are many others.  In general whenever
$\Delta_k^\Move$ is of the form $q \cdot \pdist{s_1}$ we have to
resolve the nondeterminism by choosing a $p \in [0,1]$ in
(\ref{eq:choiceofp}) above; moreover each such choice is
independent. It turns out that every extreme derivative
$\Delta'$ of $\pdist{s_0}$ is of the form
\begin{math}
  q \cdot \Lambda_0^e + (1\mathord-q)\cdot \Lambda_1^e
\end{math}
for some choice of $q \in [0,1]$, which implies that
$\Apply(\pdist{t},\pdist{q_2})$ is the convex closure of the set $\sset{\frac{1}{2}\vec{\omega},\vec{\omega}}$.
\end{example}

We have now seen two ways of associating sets of outcomes with the
application of a test to a process. The first, in
Section~\ref{sec:testing}, associates with a test and a process a set
of deterministic structures called resolutions, while the second, in
this section, uses extreme derivations in which nondeterministic
choices are resolved dynamically as the derivation proceeds.  We
proceed to show that these two approaches give rise to the same
outcomes.  The key result to this end is

\begin{proposition}\label{prop:v.der}
Let $\Lambda$ be a subdistribution in an $\omega$-respecting
deterministic pLTS $\langle R, \Omega, \rightarrow_R \rangle$. If $\Lambda\darE{}\Lambda'$ then
$\Exp{\Lambda}(\VV_{\langle R, \Omega, \rightarrow_R \rangle})=
\Exp{\Lambda'}(\VV_{\langle R, \Omega, \rightarrow_R \rangle})$.
\end{proposition}

\noindent
To obtain it, we need the crucial property that the evaluation function $\VV$ applied to $\omega$-respecting
deterministic pLTSs is continuous (with respect to  the standard Euclidean metric).

The next proposition maintains that for each extreme derivative there
is a corresponding resolution, and vice versa.
\begin{proposition}\label{prop:reso.der.1}
Let $\Delta$ be a subdistribution over the state set of a pLTS $\langle S,\Omega,\rightarrow \rangle$.
\begin{enumerate}[(i)]
\item Suppose $\Delta\darE{}\Delta'$. Then there is a resolution $\langle
R,\Lambda,\rightarrow_R\rangle$ of $\Delta$, with resolving function
$f$,   such that  $\Lambda\darE{}_R \Lambda'$ for some $\Lambda'$ for which
$\Delta' = \Imgf{\Lambda'}$.
\item Suppose $\langle R,\Lambda,\rightarrow_R\rangle$ is a resolution of a
$\Delta$ with resolving function $f$.\\ Then $\Lambda\darE{}_R\Lambda'$
  implies $\Delta \darE{} \Imgf{\Lambda'}$.
\end{enumerate}
\end{proposition}

\noindent
The definitions of outcomes, resolutions and the functional $\cal R$ directly imply that
if $\langle R, \Lambda, \rightarrow_R \rangle$ is a resolution of a
subdistribution $\Delta \inp\subdist{S}$ in a pLTS
\mbox{$\langle S, {\Omega}, \rightarrow \rangle$}, with resolving function
$f$, and $\Lambda'\in\subdist{R}$ is stable,
then $\Imgf{\Lambda'}$ is stable and\vspace{-1ex}
$$\Exp{\Lambda'}(\VV_{\langle R, \Omega, \rightarrow_R \rangle})
=\Rfun\Lambda'=\Rfun(\Imgf{\Lambda'}).$$
In combination with Propositions~\ref{prop:v.der}
and~\ref{prop:reso.der.1}, this yields:

\begin{corollary}
In an $\omega$-respecting pLTS $\langle
S,\Omega,\rightarrow\rangle$, the following statements hold.
\begin{enumerate}[(i)]
\item If $\Delta\darE{}\Delta'$ then there is a
resolution $\langle R, \Lambda, \rightarrow_R \rangle$ of $\Delta$
such that $\Exp{\Lambda}(\VV_{\langle R, \Omega, \rightarrow_R \rangle})=\Rfun(\Delta')$.

\item For any resolution
$\langle R, \Lambda, \rightarrow_R \rangle$
of $\Delta$,
 there exists an extreme
derivative $\Delta'$ such that $\Delta\darE{}\Delta'$ and
$\Exp{\Lambda}(\VV_{\langle R, \Omega, \rightarrow_R \rangle})=\Rfun(\Delta')$.
\end{enumerate}
\end{corollary}
\noindent
Together with an argument that pruning does not affect
$\RApply(\Theta,\Delta)$, this proves:

\begin{theorem}\label{cor:der.reso.2}
For any test $\Theta$ and process $\Delta$ we have that
$\Apply(\Theta,\Delta) =\RApply(\Theta,\Delta)$.
\end{theorem}

\section{Agreement of nonnegative- and real-reward must testing}\label{sec:agree}
In this section we prove the agreement of $\nrMustleq$ with $\rrMustleq$ for finitary convergent processes, by using failure simulation \cite{concur} as a stepping stone.
We start with defining the weak action relations $\dar{\alpha}$ for
$\alpha \in \Act_\tau$ and the refusal relations $\nar{A}$ for $A
\subseteq \Act$ that are the key ingredients in the definition of
the failure-simulation preorder.

\begin{definition}\rm\label{def:weak.refuse}
Let $\Delta$ and its variants be subdistributions in a pLTS $\langle S,\Act,\rightarrow \rangle$.
\begin{itemize}
\item For $a \in \Act$ write $\Delta \dar{a} \Delta'$ whenever $\Delta
  \dar{} \Delta^{\text{pre}} \ar{a} \Delta^{\text{post}} \dar{}
  \Delta'$, for some $\Delta^{\text{pre}}$ and $\Delta^{\text{post}}$. Extend this to $\Act_\tau$ by allowing as a special case
  that $\dar{\tau}$ is simply $\dar{}$, i.e.\ including identity
  (rather than requiring at least one $\ar{\tau}$).
\item For $A \subseteq \Act$ and $s \inp S$ write
  $\Ref{s}{A}$ if $s \nar{\alpha}$ for every $\alpha \inp A\cup\{\tau\}$;
  write $\Ref{\Delta}{A}$ if $\Ref{s}{A}$ for every $s \inp \support{\Delta}$.
\item More generally write $\RefE{\Delta}{A}$ if $\Delta \dar{}
  \Delta^{\text{pre}}$ for some $\Delta^{\text{pre}}$ such that
  $\Ref{\Delta^{\text{pre}}}{A}$.
\end{itemize}
\end{definition}

\begin{definition} \bf[Failure simulation preorder]\rm\label{def:efailsimA}
Define $\failsime$ to be the largest relation in $S\times\subdist{S}$
such that if $s\failsime\Delta$ then
\begin{enumerate}[(i)]
\item whenever $\pdist{s}\dar{\alpha}\Gamma'$, for $\alpha\inp\Act_\tau$, then
  there is a $\Delta' \inp\subdist{S}$ with
  $\Delta\dar{\alpha}\Delta'$ and $\Gamma'\lift{\failsime}\Delta'$,
\item and whenever $\RefE{\pdist{s}}{A}$ then $\RefE{\Delta}{A}$.
\end{enumerate}
Any relation $\mathord{\aRel} \subseteq S\times\subdist{S}$ that
satisfies the two clauses above is called a {\em failure simulation}.
The failure simulation preorder
$\mathord{\failsimpreo}\subseteq \subdist{S}\times \subdist{S}$ is defined by letting  $\Delta\failsimpreo\Gamma$
whenever there is a $\Delta\match$ with $\Delta\dar{}\Delta\match$ and
$\Gamma\lift{\failsime}\Delta\match$.
\end{definition}
Note that the simulating process, $\Delta$, occurs at the right of
$\failsime$, but at the left of $\failsimpreo$.\pagebreak[3]

\noindent
The failure simulation preorder is preserved under parallel composition
with a test, followed by pruning, and it is sound and complete for probabilistic
must testing of finitary processes.
\begin{theorem} {\bf\cite{concur}}\label{thm:sound.complete}
For finitary processes $\Delta$ and $\Gamma$,
\begin{enumerate}[(i)]
\item If $\Delta \failsimpreo \Gamma$ then for any $\Omega$-test $\Theta$ it
holds that $[\Delta\parT \Theta] \failsimpreo [\Gamma\parT \Theta]$.
\item $\Delta \failsimpreo \Gamma$ if and only if $\Delta  \pMustleq \Gamma$.
\end{enumerate}
\end{theorem}

Because we prune our pLTSs before extracting values from them, we
will be concerned mainly with $\omega$-respecting
structures. Moreover, we require the pLTSs to be \emph{convergent} in
the sense that there is no wholly divergent state $s$, i.e.\ with $s\dar{}\eDis$.

\begin{lemma}\label{lem:failsim}
Let $\Delta$ and $\Gamma$ be two subdistributions in an
$\omega$-respecting convergent pLTS $\langle
S,\Omega,\rightarrow\rangle$. If
$\Delta\failsimpreo\Gamma$, then it holds that
$\cV(\Delta)\supseteq\cV(\Gamma)$.
Here $\cV(\Delta)$ denotes $\sset{\Rfun\Delta'\mid
\Delta\darE{}\Delta'}$.
\end{lemma}

\noindent
This lemma shows that the failure-simulation preorder is a very strong
relation in the sense that if $\Delta$ is related to $\Gamma$ by
the failure-simulation preorder then the set of outcomes generated by
$\Delta$ includes the set of outcomes given by $\Gamma$. It is mainly
due to this strong requirement that we can show that the failure-simulation
preorder is sound for the real-reward must-testing preorder.
Convergence is a crucial condition in this lemma.

\begin{theorem}\label{thm:fs.rr}
For any finitary convergent processes $\Delta$ and $\Gamma$, if $\Delta\failsimpreo \Gamma$ then we have that $\Delta\rrMustleq \Gamma$.
\end{theorem}
\noindent
The proof of the above theorem is subtle. The failure-simulation preorder
is defined via weak derivations (cf. Definition~\ref{def:efailsimA}),
while the reward must-testing preorder is defined in terms of resolutions
(cf. Definition~\ref{def:d1509}). Fortunately, we have shown in
Corollary~\ref{cor:der.reso.2} that we can just as well characterise
the reward must-testing preorder in terms of weak derivations.
Based on this observation, the proof can be carried out by exploiting
Theorem~\ref{thm:sound.complete}(i) and Lemma~\ref{lem:failsim}.

This result does not extend to divergent processes. One witness example is given in Figure~\ref{fig:diverg}. A simpler example is as follows.
Let $\Delta$ be a process that diverges, by
performing a $\tau$-loop only, and let $\Gamma$ be a process that merely
performs a single action $a$. It holds that
$\Delta\failsimpreo \Gamma$ because $\Delta\dar{}{\eDis}$ and the empty
subdistribution can failure-simulate any processes. However, if we
apply the test $\pdist{t}$ from Example~\ref{ex:ex1B} again, and the
reward tuple $h$ with $h(\omega)=-1$, then
\[\begin{array}{l@{\Wide{=}}l@{\Wide{=}}l@{\Wide{=}}r}
\bigsqcap h\cdot\Apply(\pdist{t},\Delta) & \bigsqcap h\cdot\sset{\Rfun{\eDis}} &\bigsqcap \sset{0} & 0\\
\bigsqcap h\cdot\Apply(\pdist{t},\Gamma) & \bigsqcap h\cdot\sset{\vec{\omega}} &\bigsqcap \sset{-1} & -1
\end{array}\]
As $\bigsqcap h\cdot\Apply(\pdist{t},\Delta) \not\leq \bigsqcap h\cdot\Apply(\pdist{t},\Gamma)$, we see that $\Delta\not\rrMustleq \Gamma$.
Since $\cV([\pdist{t}\|\Gamma])= \{\vec{\omega}\}$ but
$\vec{\omega}\not\in\cV([\pdist{t}\|\Delta])$, this also is a
counterexample against an extension of Lemma~\ref{lem:failsim} with divergence.

Finally, by combining Theorems~\ref{thm:nr.pb}(ii) and
\ref{thm:sound.complete}(ii), together with Theorem~\ref{thm:fs.rr},
we  obtain the main result of the paper which states that, in the
absence of divergence, nonnegative-reward must testing is as discriminating as real-reward must testing.

\begin{theorem}\label{thm:rrm.nrm}
For any finitary convergent processes $\Delta$ and $\Gamma$,
it holds that $\Delta\rrMustleq \Gamma$ if and only if
$\Delta\nrMustleq \Gamma$.
\end{theorem}

\section{Conclusion}\label{sec:conclusion}
We have studied a notion of real-reward testing which extends the
traditional nonnegative-reward testing with negative rewards. It
turned out that real-reward may preorder is the inverse of real-reward
must preorder, and vice versa. More interestingly, for finitary
convergent processes, the real-reward must testing preorder coincides with
the nonnegative-reward testing preorder. In order to prove this result, we
have presented two testing approaches and shown their coincidence,
which involved proving some analytic properties such as
the continuity of a function for calculating testing outcomes.

Although for finitary convergent processes real-reward must testing
is no more powerful than non\-negative-reward must testing,
the same does not hold for may testing. This is immediate from
our result that (the inverse of) real-reward may testing is as
powerful as real-reward must testing, that is known not to hold for
nonnegative-reward may- and must testing.
Thus, real-reward may testing is strictly more discriminating than
nonnegative-reward may testing, even without divergence.

\bibliographystyle{eptcs}
\bibliography{prob}
\end{document}